\documentclass[preprint,pre,aps,superscriptaddress,a4paper]{revtex4}
\usepackage[utf8x]{inputenc}
\usepackage{graphicx}
\usepackage{amsmath}
\usepackage{amsfonts}
\usepackage{color}
\usepackage{tabularx}
\usepackage{textcomp}

\begin{document}
\title{Shear induced ordering in systems with competing interactions: A machine learning study}
\author{J. P\c ekalski}
\email{jpekalski@ichf.edu.pl}
\affiliation{Institute of Physical Chemistry, Polish Academy of Sciences, Kasprzaka 44/52, 01-224 Warszawa, Poland}

\affiliation{Department of Chemical and Biological Engineering, Princeton University, Princeton, New Jersey 08544, USA}

\author{W. Rz\c adkowski}
\affiliation{Institute of Science and Technology Austria (IST Austria), Am Campus 1, 3400 Klosterneuburg, Austria}
\author{A. Z. Panagiotopoulos}
\affiliation{Department of Chemical and Biological Engineering, Princeton University, Princeton, New Jersey 08544, USA}
%\date{\today}
\begin{abstract}
When short-range attractions are combined with long-range repulsions in colloidal particle systems, complex microphases can emerge. Here, we study a system of isotropic particles which can form lamellar structures or a disordered fluid phase when temperature is varied. We show that at equilibrium the lamellar structure crystallizes, while out of equilibrium the system forms a variety of structures at different shear rates and temperatures above melting. The shear-induced ordering is analyzed by means of principal component analysis and artificial neural networks, which are applied to data of reduced dimensionality. Our results reveal the possibility of inducing ordering by shear, potentially providing a feasible route to the fabrication of ordered lamellar structures from isotropic particles.

\end{abstract}
\maketitle

\section{Introduction}
When charged spherical hard-core particles attract each other at short distances, their pair interactions can take the form of the short-range attraction, long-range repulsion (SALR) potential. Such competing interactions were shown to appear between colloidal particles \cite{ghezzi:97:0} or globular proteins \cite{stradner:04:0} with a proper balance between the electrostatic and solvent-mediated forces. Theory and computer simulations predict SALR particles to self-assemble into clusters and elongated aggregates that form either disordered glassy and gel states \cite{sciortino:05:0,toledano:09:0} or periodically ordered morphologies such as fcc ordered clusters, hexagonally ordered columns, double gyroid structure or the lamellar phase \cite{zhuang:16:0,zhuang:16:1,zhuang:16:2,ciach:10:1, archer:08:0,ciach:08:1,candia:06:0}. Experimentally, however, the ordered structures are observed only for SALR particles adsorbed at two-dimensional (2d) interfaces \cite{ghezzi:97:0,law:13:0,sear:99:0}, while in a three-dimensional (3d) bulk the efforts to form ordered structures have been so far unsuccessful \cite{royall:18:0}. Although the realization of the SALR potential in 3d real space is quite challenging, formation of 3d ordered structures by isotropic particles is of significant interest.

One way to gain stability enhancement of an ordered structure is to confine the system with walls whose geometry fits the symmetries of the ordered pattern. In the case of the SALR system it was shown that as long as the distance between two parallel walls fits to the period of the stripe structure, the stability of the ordered phase is enhanced \cite{almarza:16:0,pekalski:19:0}. Shorter wall-wall separation corresponds to a higher melting temperature according to a Kelvin-like equation \cite{almarza:16:0}. Thus, confinement can effectively suppress thermally induced topological defects, but an accurate choice of the distance between the walls is hard to realize.
 
Another possible method to induce ordering is to apply shear. {\color{black} In the case of homogeneous systems shear was shown to either suppress~\cite{palberg:95:0,blaak:04:0} or enhance crystallization~\cite{ackerson:88:0,amos:00:0} depending on system and conditions~\cite{catherall:00:0}. Inhomogeneous systems, such as liquid crystals~\cite{li:04:0} or diblock copolymers~\cite{harrison:20:0,nikoubashman:14:0}, when exposed to steady shear 
were shown to order into layered structures. The focus of the present study is on whether also the inhomogeneous SALR fluid orders when exposed to steady shear.} In the case of SALR systems, the first reports on shear effects were provided by Imperio \cite{imperio:08:0}, but only ordered monolayers were sheared. In 3d, the shear effects on SALR systems have been studied only recently \cite{ruiz:19:0,stopper:18:0}. In Ref. \cite{ruiz:19:0} the authors showed that the equilibrium gel structure after being exposed to steady shear exhibits local, short length-scale anisotropies; however no global ordering was found. In Ref. \cite{stopper:18:0}, by means of the classical density functional theory, the effects of shear on ordered states were described, showing e.g. shear induced transition between the double gyroid and the cylindrical phases. Here, we describe a 3d SALR system that forms an ordered structure when exposed to steady shear, even above the melting temperature of the ordered phase.
 
Although SALR systems can spontaneously form a variety of microphases~\cite{ciach:13:0,ciach:08:1}, we focus on the lamellar phase only. The equilibrium properties of the SALR lamellar phase were intensively studied with a variety of methods at different approximation levels. In 2d bulk the lamellar phase consists of parallel stripes that were shown to melt in a  step-wise manner \cite{almarza:14:0}. First, the translationally ordered low temperature structure undergoes a transition into a molten lamella phase with only orientational ordering. Further heating results in a transition into a disordered, yet inhomogeneous fluid. This finding from computer simulations agrees with the mean-field calculation obtained in a 1d approximation of the lamellar system that shows coexistence of two ordered phases characterized by different amplitudes of the density profiles~\cite{pekalski:13:0}. In 3d however, no signs of the step-wise melting of the lamellar phase were reported \cite{zhuang:16:0,imperio:04:0,archer:08:0,candia:06:0}. We will show here that also in 3d SALR particles can spontaneously form two distinct structures that are both lamellar.

Another question that will be addressed in this work, is how far the level of description complexity can be reduced so that different 3d structures can still be distinguished and transitions between them localized. To do that, instead of analyzing 3d particle configurations, we will use binary matrices obtained by mapping the 3d structure onto a 2d discretized surface. The maps will be then utilized as input data for different machine-learning based methods aimed at pattern recognition.

{\color{black} Machine learning~\cite{carleo:19:0, mehta:19:0}, in particular methods based on artificial neural networks~\cite{carrasquilla:17:0}, has been recently proven successful in structure recognition for a wide range of physical systems. The general ability to identify arbitrary patterns in data makes these methods especially appealing when the number and/or features of interest are not known. In machine learning, there are two main approaches: supervised learning and unsupervised learning. In the supervised learning paradigm, at first labelled data is fed into the algorithm. Based on that, training data patterns are inferred and can be later used to classify new examples of data. In the unsupervised paradigm, different phases are identified without any labeled data. In this work, we employ both approaches. We use principal component analysis (PCA) for unsupervised classification and artificial neural networks as a supervised approach. We combine the two approaches and verify their usability for pattern recognition by comparing with a standard analysis of the structure factor.

}

The manuscript is organized as follows: in Sec. \ref{sec:mm} the SALR potential is defined and the methods used for production and analysis of the equilibrium and nonequilibrium structures are described. In Sec. \ref{sec:results} the results are presented, and in Sec. \ref{sec:summary} we summarize and discuss them in context of previous research. 

\section{Model and methods}
\label{sec:mm}

% Model
\subsection{Model}
\label{subsec:model}

\begin{figure}[ht]
    \begin{center}
    \includegraphics[scale=1]{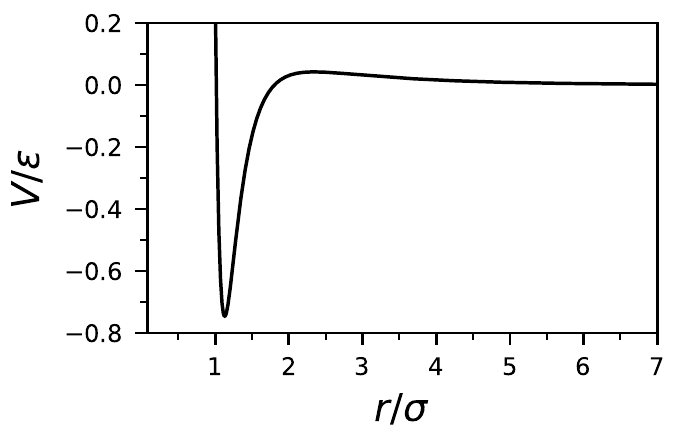}
      \caption{The SALR pair potential. It crosses zero at $r = 1.014\sigma$, then it has a local minimum for $r = 1.13 \sigma$, crosses zero again at $r =1.81$ and has a local maximum for $r = 2.34  \sigma$.
      The potential has been cut off at $r= 7\sigma$.
      }
      \label{fig:pot}
    \end{center}
  \end{figure}

We used the SALR potential of the form previously studied in Refs.\cite{sciortino:04:0,sciortino:05:0,mani:14:0,toledano:09:0,santos:17:0}, where effective interparticle interaction is the sum of the standard Lennard-Jones and Yukawa potentials:
\begin{equation}
V(r) = 4 \varepsilon \left[\left( \frac{\sigma}{r} \right) ^{12} - \left( \frac{\sigma}{r} \right) ^6 \right] + \frac{A}{r} e ^{-r \kappa}.
\label{Vr}
\end{equation}
 The parameter values were taken from \cite{santos:17:0}: $A = 0.5$, $\kappa = 0.5$, $\varepsilon = 1.0$, $\sigma =1.0$, where $\varepsilon$ and $\sigma$ are set to be the units of energy and length respectively (Fig. \ref{fig:pot}). Accordingly, in reduced units temperature is $k_BT/\varepsilon$, where $k_B$ is the Boltzmann constant, and time is given in units of $\tau = \sqrt{\frac{m \sigma^2}{\varepsilon}}$, where $m=1$ is the mass scale. For this set of parameters, particle aggregation into clusters with preferred size occurs up to $k_BT/\varepsilon \approx 0.45$, even though the potential is found to be repulsion-dominated (positive second virial coefficient) for $k_BT/\varepsilon > 0.31$ \cite{santos:17:0}. {\color{black} Apart from a detailed description of the low density behavior, for the chosen set of parameters the phase behavior of the system is not known precisely, but can be inferred by analogy to similar systems \cite{zhuang:16:0,ciach:08:1, ciach:13:0,candia:06:0}.}

\subsection{Simulation details}
\label{sec:sim}
% simulation method
 
 In order to minimize size effects and to ensure commensurability between box size and the period of the formed structures the molecular dynamic simulations (MD) were performed in the isobaric-isothermal ensemble, {\it NPT}, where $N$ is the number of particles, $P$ is the pressure and $T$ is the temperature. We used $N=15625$ and $P = 0.5 \varepsilon/ \sigma^3$.
 
 Equilibrium simulations were run using the HOOMD-blue package~\cite{hoomd:2,hoomd:1} with a time step $dt = 0.005 \tau$. The runs consist of $10^8$ MD steps with either linearly increasing or decreasing temperature in the range of $0.01<k_BT/\varepsilon<0.8$. For decreasing temperature runs the initial configuration was a simple cubic lattice, and the final configuration of this procedure was later used to initialize the linear heating protocol. {\color{black} For temperatures away from the phase transitions both heating and cooling protocols gave the same number densities of the system, thus we can infer that the equilibrium was reached.}
 
 The nonequilibrium MD simulations were performed in LAMMPS~\cite{lammps:1} adjusted to use the SLLOD equation of motion in the NPT ensemble~\cite{sllod:book}. The shearing procedure was initialized from the equilibrium structures, then after the $10^6$ steps needed to reach the stationary state, production runs consisting of $10^6$ steps were executed. The shear was imposed in the $x$ direction, and the velocity gradient was in the $y$ direction. Therefore, the shear rate is defined as $R= v_x/y$ and is measured in inverse time units, $\tau^{-1}$.
 
\subsection{Analysis details}
\label{sec:anal}

Dynamical properties of the equilibrium structures were described by computing components of the average displacement vector, $D = (d_x,d_y,d_z)$, given by
\begin{equation}
    d_k = \frac{1}{N}\sum_{i=1}^N [R_i(t_0 + \Delta t)-R_i(t_0)], \quad k = x,y,z, 
\end{equation}
where $R_i(t)$ is the $i$-th particle location at time $t$, and $\Delta t = 10^4  dt$. The calculations were performed after rotating the structure so that $d_{\parallel} = (d_x+d_y)/2$ and $d_{\perp} = d_y$ describe the displacement in directions parallel and perpendicular to the plane of the slabs respectively. For the temperature dependence analysis, time averages of $d_k$ were calculated over $10^2$ consecutive configurations of the cooling simulation protocol.

Structural analysis of the sheared fluid was performed using three methods: principal component analysis (PCA), dense neural networks (DNN) and convolutional neural networks (CNN). The motivation to compare these methods comes from their potential to combine the power of unsupervised and supervised learning. 

\begin{figure}[ht]
    \begin{center}
    \includegraphics[width=0.5\textwidth]{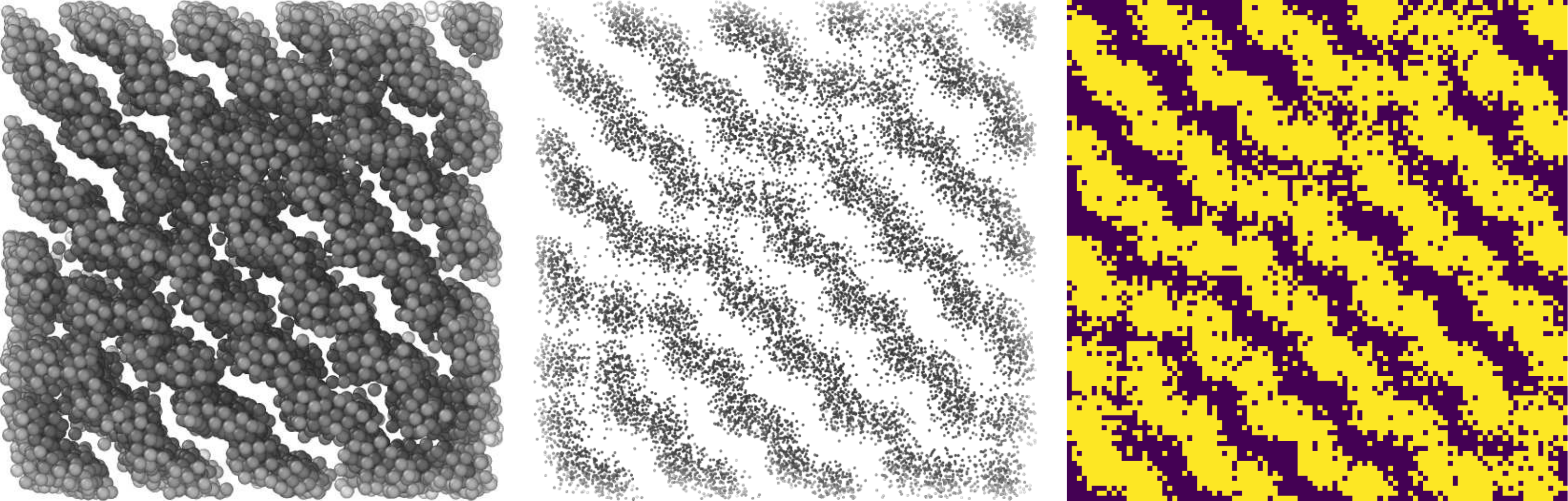}
      \caption{Stages of binary map production. Left panel: processed particle configuration seen from the top view of the (y,z) plane, perpendicular to the shear flow. Middle panel: spherical particles transformed to point particles. Right panel: (y,z) positions of the point particles were binned into 100x100 grid. The grid was then binarized depending on whether the bin was occupied (yellow color) or not not (dark color).
      }
      \label{fig:mapping}
    \end{center}
\end{figure}

All the algorithms aimed at structure analysis were fed with the same data, that is a 2d map of the 3d structure obtained by perpendicular projection of particle positions on a $100\times100$ grid located at the $x=0$ plane, i.e. the plane perpendicular to the flow direction. The normalized map was then transformed to a binary array, $M$, by the ceiling function which turns every nonzero value to 1. As a result, the only information contained in each element of $M$ is whether at a given time-step there is a particle with $y$ and $z$ coordinates within a grid node regardless of its $x$ coordinate value. An example of how a lamellar structure is projected on 2d binary matrix is shown in Fig. \ref{fig:mapping}.

The PCA method~\cite{jolliffe:16:0} is an unsupervised learning method often utilized for pattern recognition. It involves calculating eigenvalues, $\lambda_i$, usually via the singular value decomposition of the matrix that is symmetric and positive semi-definite. In our case, for the location of structural transitions we used $\lambda_1$, the maximal eigenvalue of the $M \times M^T $ matrix. To compare $\lambda_1$ between different temperatures we normalize it by its maximal value at each temperature: $\lambda \equiv  \lambda_1 / \max_{R} \lambda_1(R) $. {\color{black}Let us also note that other, more involved unsupervised machine learning methods~\cite{wetzel:17:0}, such as support vector machines~\cite{cubuk:15:0, schoenholz:16:0, schoenholz:17:0} or variational autoencoders~\cite{cristoforetti:17:0}, have been proven successful in phase classification of physical systems. However, we employ the PCA method as we focus on the lamellar structure, whose ordering should be simple enough to be captured even by this more elementary tool.}

Artificial neural networks{\color{black}~\cite{graupe:13:0}} are commonly used for data classification based on features that they were trained to distinguish. {\color{black} They are a powerful tool for encoding functions that describe a set of features (e.g. Monte Carlo simulation, image data) into a set of classes (e.g. ordered and disordered phase). This encoding is achieved by decomposing the function into a set of small units, called neurons. The behavior of individual neurons depends on a set of trainable parameters that are gradually adjusted during the training phase. The training involves minimization of a cost function that quantifies the difference between the ground truth labels and current neural network output over the set of the neural network parameters. } 
CNNs are an example of networks that are designed for pattern recognition based on the translationally invariant two-dimensional features of their inputs~\cite{gu:18:0, dhillon:19:0}. On the contrary, DNNs do not convolve the input with a set of filters; the input is instead densely connected to the first hidden layer of neurons. Hence, in that case the preparation of data for training includes flattening, a transformation from 2d matrices to 1d vectors. Thus in the case of DNN, the learning relies on finding correlations between artificially produced vectors. We have used both types of networks to categorize the binary maps, $M$. For both networks we used maps obtained at $k_BT/\varepsilon = 0.42$ with $R\tau= 0.01,0.1,30$ (low and high shear disordered structure), $R\tau= 6,8,10$ (ordered lamella) and $R\tau= 16,20,24$ (torn lamella). For each value of $R$ we used $400$ maps for training and the rest (100 maps) for validation. The DNN features a multilayer perceptron architecture with the input layer being followed by a 32-neuron hidden layer activated with ReLU activation~\cite{nair:10:0}. The output layer is made of three softmax-axtivated neurons. The CNN consists of two convolutional layers followed by a softmax-activated output layer. The convolutional layers feature 32 (input) and 16 (hidden) filters of size 3x3.  Both networks were trained using the Adam algorithm with the aim to minimize the cross entropy between the softmax output and ground truth labels. Despite the fairly simple and shallow architecture, for both sets we reached a classification accuracy of 1.0 and cross entropy loss function values of 0.0008 for the training set and 0.005 for the validation set.

{\color{black} The validity of the machine learning approach was verified by computing an order parameter, $O_p$, based on the structure factor, $S(k)$. To find the structure factor the following formula was used
\begin{equation}
    S(k) = \frac{1}{N} \left[ \sum_{i=1}^N \sin (k \cdot R_i)  \right]^2 + \frac{1}{N} \left[ \sum_{i=1}^N \cos (k \cdot R_i)  \right]^2,
\end{equation}
where $\cdot$ is the scalar product, and to describe the shear-induced lamellar ordering the wave vector, $k$, was set to point in the direction perpendicular to the lamellar slabs. Usually, the order parameter is the normalized height of the highest $S(k)$ peak that appears for $k>2\pi/L$, where $L$ is the size of the simulation box. Here, in order to fit $O_p$ to the scale of $\lambda$, we transform $S(k_{max})$ linearly in the following way: $O_p = \frac{1}{2}(1- S(k_{max}))+\lambda_{min}(R)$, where $\lambda_{min}(R)$ is the minimal value of $\lambda(R)$ at given temperature.

}

The visualization of the snapshots was produced using The Open Visualization Tool (OVITO) \cite{ovito}.

\section{Results}
\label{sec:results}

\subsection{The Equilibrium {\color{black}Case}}

\begin{figure}[ht]
    \begin{center}
    \includegraphics[scale=1]{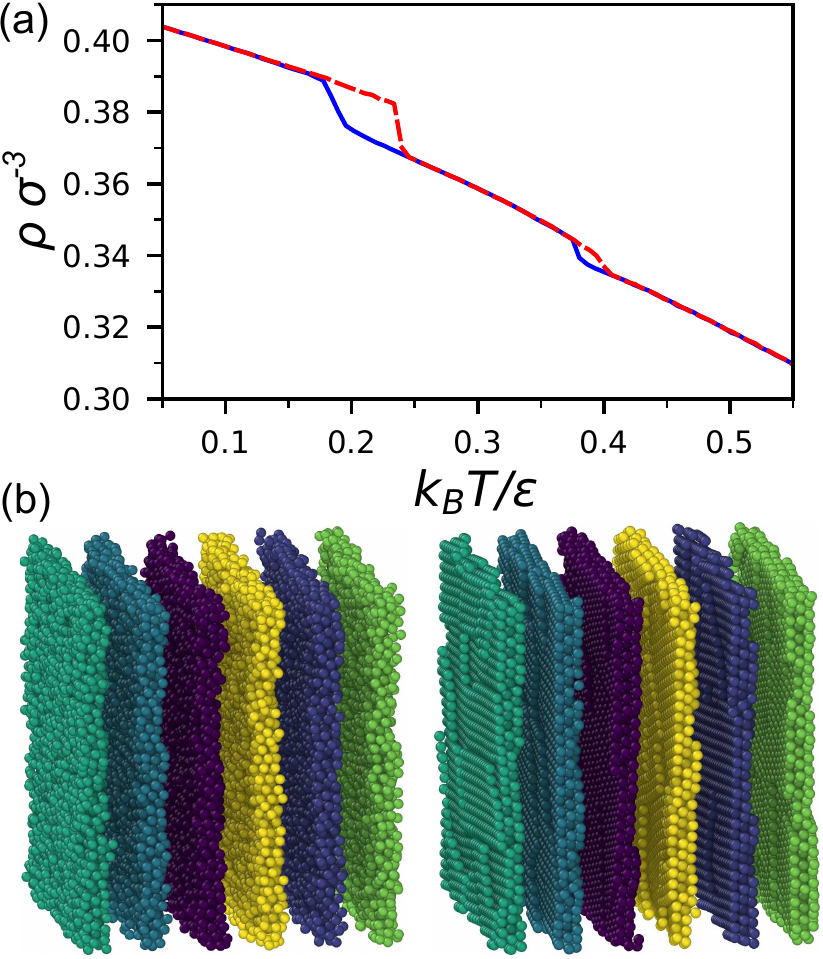}\\
      \caption{(a) Number density as a function of temperature at constant pressure $P = 0.5 \varepsilon/ \sigma^3$ for a sequence of simulations with linearly increasing (dashed red lined) or decreasing (solid blue line) temperature. The density drops appear at $k_BT/\varepsilon= 0.185$ and  $0.377$ for decreasing $k_BT/\varepsilon$, while at $k_BT/\varepsilon= 0.237$ and $0.40$ for increasing. (b) Representative snapshots of lamellar structures with fluid-like (left) and crystal-like (right) ordering.
      }
      \label{fig:vol}
    \end{center}
\end{figure}

SALR particles form ordered microphases only for a specific range of model parameters which provides a proper balance between the attractive and repulsive interactions~\cite{zhuang:16:2}. In the case of very narrow attraction wells, aggregates are formed, but global ordering does not occur~\cite{sciortino:05:0,toledano:09:0,ruiz:19:0}. Here, following \cite{santos:17:0,archer:08:0,zhuang:16:0} we choose a SALR potential with a wider attraction well. We have checked that the chosen parameters lead to self-assembly into globally ordered cluster, columnar and lamellar structures. Interestingly, the ordering occurs also at temperatures for which the SALR potential has a positive second virial coefficient~\cite{santos:17:0}. In Fig.~\ref{fig:vol} the lamellar structures obtained from equilibrium MD simulations in the NPT ensemble are presented. The phase transition between the lamellar structures with crystal-like slabs and fluid-like slabs is reflected in a drop of the average density. The transitions occur at different temperatures depending on the simulation protocols used, as described in the caption of Fig. \ref{fig:vol}. {\color{black}A similar but larger hysteresis is observed for the high-temperature melting transition. In both cases the observed density response suggests presence of a first order transition. The melting of the crystal-like ordering preserves lamellar structure, occurs only within slabs, and is a local change that leads to a relatively small hysteresis. On the other hand, in the high-temperature transition the melting turns a lamellar structure into a disordered fluid, and such global order-disorder transition results in a relatively large hysteresis.} Transition between lamellar structures with different slab characteristics was not previously reported for 3d systems. However in 2d, similar multi-step melting of stripes was found~\cite{Reichhardt2010}. 

\begin{figure}[ht]
    \begin{center}
    \includegraphics[scale=1]{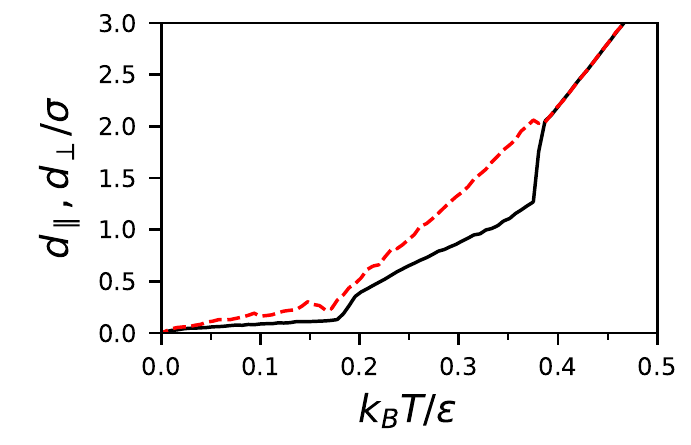}
      \caption{Average displacements versus temperature in directions parallel (dashed red) and perpendicular (solid black) to the lamellar slabs.
      }
      \label{fig:disp}
    \end{center}
\end{figure}

The analogy with the 2d system is not only structural but also dynamical, since the displacements of the particles along different axes behave in a similar manner as in the 2d system (Fig. \ref{fig:disp} and Fig. 8 in \cite{Reichhardt2010}). In particular, formation of the lamellar structure causes a significant drop in the average particle displacement along the axis perpendicular the slabs, $d_{\perp}$, and simultaneously a slight increase of $d_{\parallel}$. The difference between $d_{\perp}$ and $d_{\parallel}$ reflects the anisotropy of the system and the suppression of particle exchange between the slabs. Upon further decrease of temperature both the anisotropy and the displacements converge and the crystal lamellar structure is formed.

\subsection{Nonequilibrium Simulations} 

\begin{figure}[ht]
    \begin{center}
    \includegraphics[width=0.2\textwidth]{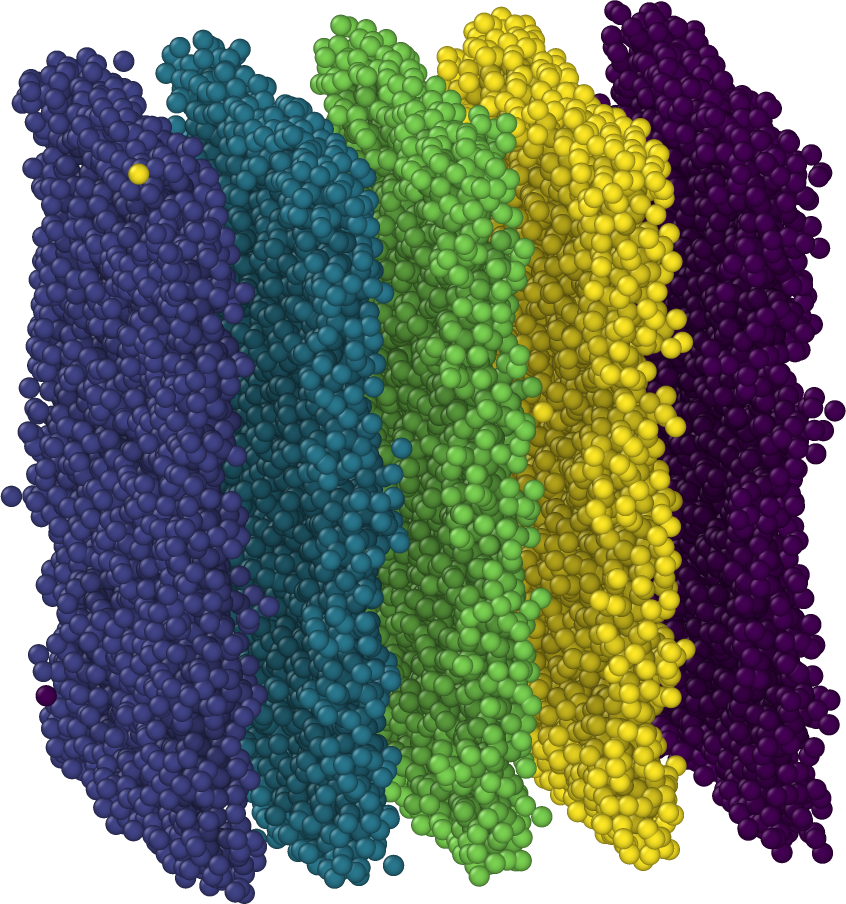}
    \includegraphics[width=0.2\textwidth]{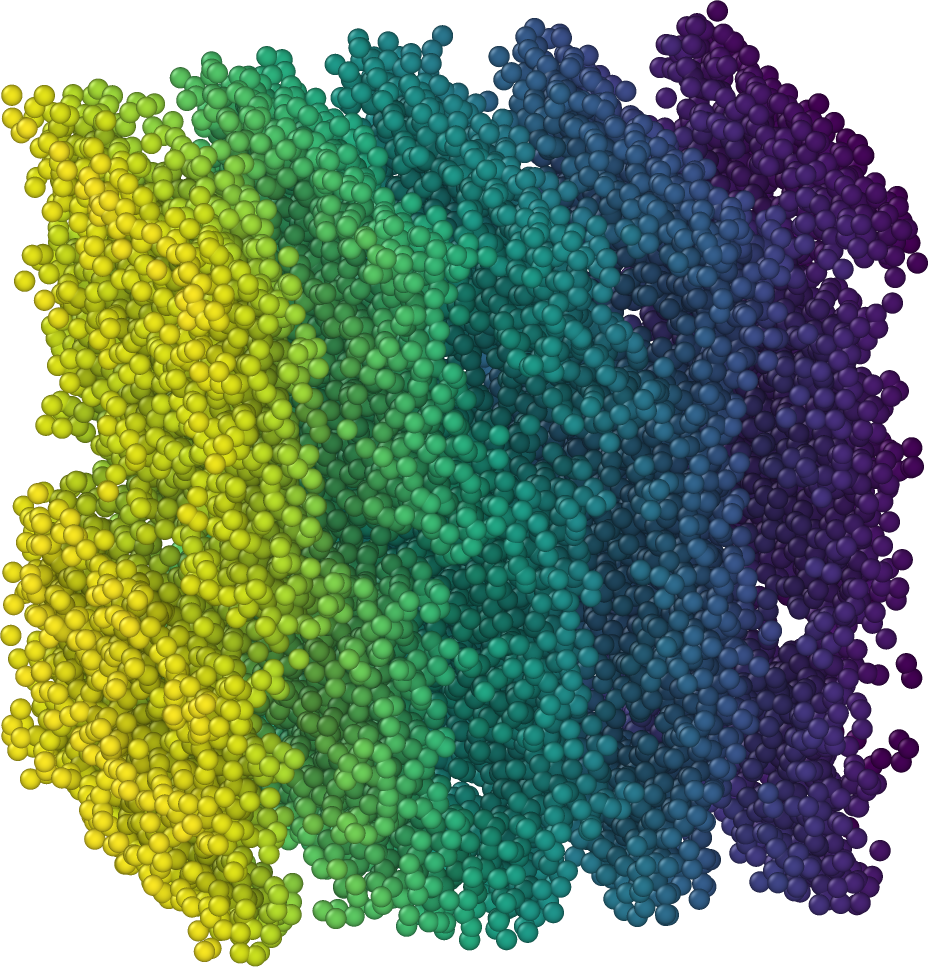}\\    
      \caption{Representative steady states obtained after exposing disordered fluid to shear. Left panel: $(k_BT/\varepsilon,R\tau) = (0.42, 10)$, right panel: $(k_BT/\varepsilon,R\tau) = (0.65, 10)$.}
      \label{fig:snaps_shear}
    \end{center}
\end{figure}

In the case of a SALR potential that leads to periodically ordered structures, the effect of shear on the ordered microphases was studied in Ref. \cite{stopper:18:0}. Thus, here we focus on the question of whether steady shear can induce ordering of disordered SALR fluids. We apply shear to the equilibrium structures at $P=0.5\varepsilon/ \sigma^3$ and for $k_BT/\varepsilon>0.4$, that is, above the melting temperature of the lamellar phase. We find that the obtained stationary states show different kinds of ordering depending on the applied shear-rate and the temperature. Importantly, close to the equilibrium order-disorder phase transition, the steady state structures (Fig. \ref{fig:snaps_shear}) are highly similar to the equilibrium lamella with fluid-like slabs (Fig. \ref{fig:vol}). However, since many other states were found that were not as highly ordered, in what follows we describe and analyze structural transitions between different shear-induced morphologies by applying different machine learning methods to 2d binary maps of the obtained 3d structures, as described in Sec. \ref{sec:anal}. 
 
\subsection{Principal Component Analysis}

\begin{figure}[ht]
    \begin{center}
    \includegraphics[scale=1]{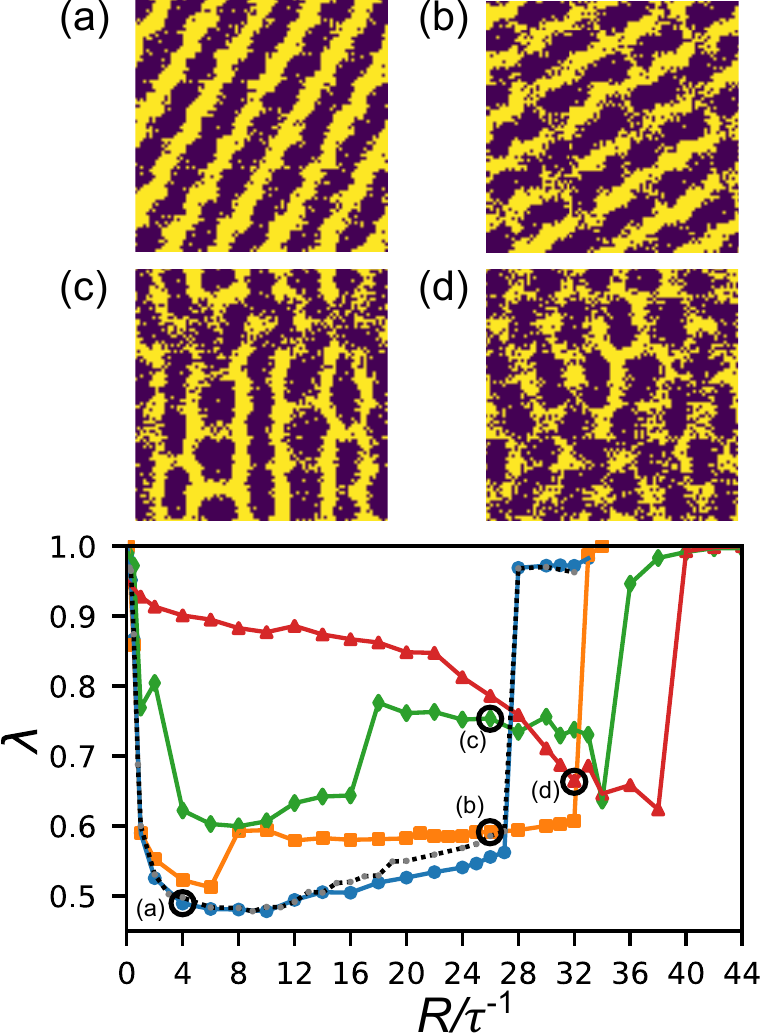}\\
      \caption{Normalized principal component, $\lambda$, versus shear rate, $R$, at different temperatures: $k_BT/\varepsilon = 0.42$ (blue circles), $k_BT/\varepsilon = 0.55$ (orange squares), $k_BT/\varepsilon = 0.65$ (green diamonds), $k_BT/\varepsilon = 0.75$ (red triangles). {\color{black} The black dotted line is the structure factor order parameter, $O_p$, calculated for $k_BT/\varepsilon = 0.42$.} The upper panels show representative structures observed at different $\lambda$ values marked by an open circle and corresponding to (a) $(k_BT/\varepsilon, R\tau) = (0.42, 4)$, (b) $(0.55,26)$, (c) $(0.65,26)$ and (d)  $(0.75,32)$. 
      }
      \label{fig:pca}
    \end{center}
\end{figure}

The PCA results shown in Fig. \ref{fig:pca} indicate the presence of five distinct morphologies, with different levels of ordering described by the value of $\lambda$.  In the low and high shear limits $\lambda=1$, which corresponds to disordered steady states. In the mid range of shear-rates $\lambda$ drops significantly, and the ordered structures appear. At $k_BT/\varepsilon = 0.42$, the shear induced structure is the lamella with fluid-like slabs (Fig.\ref{fig:pca}a). The optimal shear-rate for this conformation is when $\lambda(R)$ is at the minimum, that is for $R\tau= 10$. At $k_BT/\varepsilon = 0.42$,  with an increasing shear rate the $\lambda$ value gradually increases, and defects in the lamellar structure start to appear. The high shear tears the lamellar slabs, so that their integrity is broken and some slabs connect (like in Fig. \ref{fig:pca}b). However, the global ordering persists up to $R\tau=26$ where a structural transition to a disordered fluid occurs. The number of topological defects present in the torn lamellae varies with shear rate and temperature. The torn lamellae occur when $\lambda \approx 0.6$, and thus are present also at $k_BT/\varepsilon = 0.55$  (for $8\le R\tau\le32$) and $k_BT/\varepsilon = 0.65$ (for $4\le R\tau\le 16$). With higher temperatures, less ordered structures form. At $k_BT/\varepsilon = 0.65$ for $18\le R\tau\le 32$, the structure is no longer globally lamellar, as slabs appear along with columns (Fig. \ref{fig:pca}c). At $k_BT/\varepsilon = 0.75$, temperature fluctuations destroy lamellar order entirely and the drop of $\lambda$ with increasing shear is less abrupt. When $\lambda$ gradually decreases towards the value of $0.65$, a weak ordering occurs and the structure resembles hexagonally ordered columns (Fig. \ref{fig:pca}d).

\subsection{Artificial Neural Networks}
 
 The PCA method reduces the dimensionality of the data that it was fed with. Here, PCA was used to analyze data already reduced to 2d discrete space. Thus, the question is now if we did not lose significant information along the way. To verify this, we will analyze structural behavior of the sheared system with more complex, neural network based methods.
 
 We applied two commonly used supervised methods for pattern recognition, that is DNN and CNN. The aim of the training was to  learn to classify the structures that occur at $k_BT/\varepsilon = 0.42$, where PCA suggests that in between critical shear rates required for ordering, a gradual structural transition from ordered to torn lamella is present. The results are presented in Fig. \ref{fig:nn}. In both cases the steady states for $R\tau\leq 0.5$ and $R\tau\ge 28 $ are labeled by the networks as disordered, which agrees with the PCA predictions. For the mid range of the shear rate, both networks are able to distinguish the lamellar and the torn lamellar structures and predict that the transition occurs between $11 < R\tau< 20 $. This suggests that in contrast to the order-disorder transitions these steady states change gradually. Thus, both neural networks' predictions seem to agree with the results from the PCA analysis. Interestingly, this is not the case for $k_BT/\varepsilon = 0.55$ (not shown), where the thermal fluctuations are more prominent. The structures are far less ordered, and only CNN predictions to some extent agree with PCA whereas the DNN does not even capture the location of the transitions between disordered and the torn-lamellar structures. One should note however, that both networks were trained with data from $k_BT/\varepsilon = 0.42$.
 
 \begin{figure}[ht]
    \begin{center}
    \includegraphics[width=0.5\textwidth]{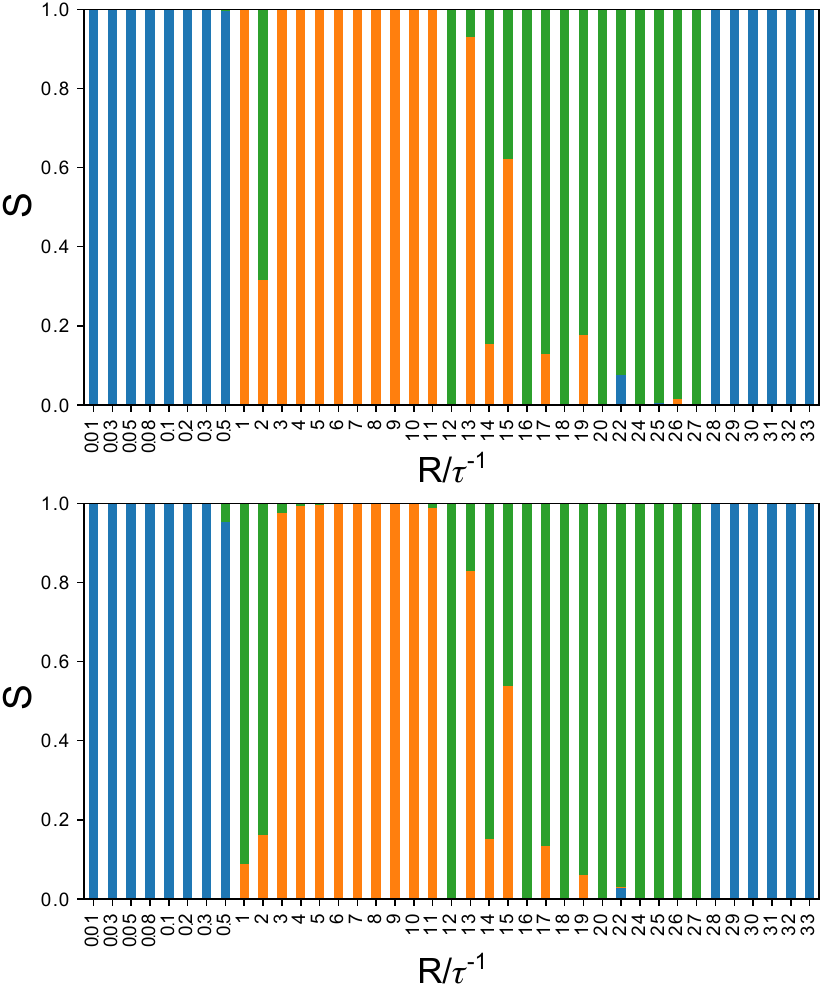}
      \caption{Fraction of configurations assigned to different structures by DNN (upper panel) and CNN (lower panel) as a function of the shear rate at $k_BT/\varepsilon = 0.42$. The neural networks were trained to distinguish between the disordered structures (blue), lamellar structure (orange) and torn lamellar structure (green).
      }
      \label{fig:act}
    \end{center}
\end{figure}

 It is of interest to learn how the neural network makes the final decision on how a given structure is labelled. Some insight into that process can be found from the outputs of the last layer of the CNN. The number of neurons in the output layer corresponds to the number of structures that the network was trained to distinguish. The output values $p_1,p_2$ and $p_3$ corresponds to the lamellar, torn lamellar and disordered structures respectively and take values in the range $[0,1]$. The final classification decision is made by choosing the structure corresponding to the largest-output softmax neuron of the output layer. In Fig. \ref{fig:act}, the output value $p_1$ is shown for the shear rates at which the PCA predicts gradual transition between the lamellar and the torn lamellar structures. In this range of shear rates $p_2 = 1-p_1$ and $p_3 = 0$. The ongoing structural transition is reflected in the high noisiness of the output. In particular, at the shear rate of $R\tau=13$ where more then $90\% $ of configurations were labelled as lamellar, the output plot shows that actually in many cases the $p_1 \approx 0.5$, meaning the output values of the lamellar and torn structures are close. This suggests that the network, although with high accuracy, made the decision with less certainty. An increase in the shear rate makes the predictions more ambiguous, so that for $R\tau= 15$, where about half of the configurations are labelled as lamellar and half as torn lamellar, the output resembles random noise. Increasing the shear rate to $R\tau=17$ changes the situation dramatically. For most of the configurations, the classification decisions are made with high certainty. However, for some range of configurations, a kind of transition between the steady states can be observed. Further shear rate increases result in monostability, and the outputs become constants of value 0 or 1 (not shown).

\begin{figure}[ht]
    \begin{center}
    \includegraphics[width=0.5\textwidth]{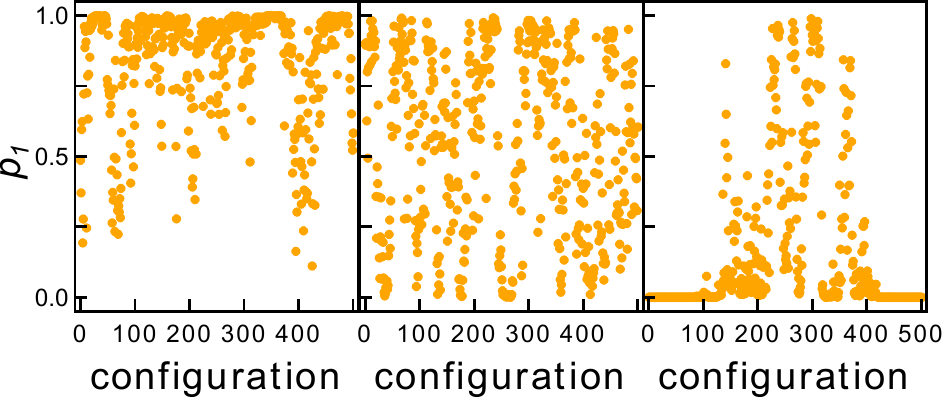}
      \caption{Output, $p_1$, of neuron corresponding to the lamellar structure in the last, softmax-activated layer of the convolutional neural network upon the gradual transition from the lamellar to torn lamellar structures at $k_BT/\varepsilon = 0.42$. From left to right: $R\tau= 13, 15, 17$.}
      \label{fig:nn}
    \end{center}
\end{figure}

{\color{black}
\subsection{Structure Factor}
Finally, we would like to compare the proposed methodology with the standard approach that is based on the structure factor analysis. In Fig.~\ref{fig:pca}, along with the PCA results we present a one-dimensional structure factor order parameter, $O_p$, that was calculated based on non-reduced particle coordinates as described in Sec. \ref{sec:mm}. Both $\lambda$ and $O_p$ indicate the same values of the critical shear rates corresponding to the structural transitions between the disordered and ordered steady states. Moreover, $O_p$ takes the same values as $\lambda$ for $R\tau\le 14$, where, as found with neural networks analysis, most of the configurations are still lamellar with no defects. When the torn lamellar structures take over, a systematic difference in the values of $O_p$ and $\lambda$ can be observed. The reason is that the torn lamellae are not periodic and a one-dimensional description is no longer sufficient. We also found that the $O_p(R)$ at higher temperatures, where lamellar structures are no longer present, exhibits nonmonotonical behavior as a function of $R$ and thus does not correctly describe the ongoing ordering (not shown).

}

\section{Discussion and conclusions}
\label{sec:summary}

We have analyzed properties of lamellar structures formed by a SALR fluid in equilibrium and nonequilibrium conditions. We have found that the equilibrium lamellar phase can be formed with slabs that exhibit either fluid- or crystal-like ordering. The transition between the two is manifested by a jump in the density and by the average displacement becoming significantly anisotropic upon heating. {\color{black} Although the presence of crystal-like lamellar phase with hexagonally packed particles has been reported before~\cite{candia:06:0}, the phase transition between the two ordered lamellar phases has not been shown in} previous attempts to describe phase behavior of the 3d SALR fluid~\cite{zhuang:16:0,imperio:04:0,archer:08:0,candia:06:0}. Interestingly, two distinct lamellar phases have been found in 2d lattice~\cite{almarza:14:0} and off-lattice~\cite{Reichhardt2010} models, as well as in the model for a 1d cross-section of the stripe phase \cite{pekalski:13:0}. Importantly, the nature of the transition found here and in Ref. \cite{Reichhardt2010} is the same and shows a liquid-crystal transition within the slabs. On the other hand, the molten lamellar phase described in \cite{almarza:14:0}, consists of liquid-like stripes with randomly distributed topological defects.  In the 3d equilibrium, no such defects were observed, but this can be model-dependent and requires further studies.

Out of equilibrium, the disordered SALR fluid was shown to form a number of anisotropic morphologies when exposed to steady shear. In particular, depending on temperature and applied shear rate, we have found that lamellar, torn lamellar, lamellar mixed with columnar, and hexagonally-ordered columnar structures can occur. The anisotropic structures with weak, global ordering were observed up to $k_BT/\varepsilon = 0.75$, with $k_BT/\varepsilon = 0.4$ being the temperature of the equilibrium order-disorder phase transition.

At equilibrium, one of the main obstacles that a system with competing interactions has to overcome to form an ordered state is finding the global minimum in a highly complex energy landscape. The barriers between the local minima that potentially can trap the system are steeper at lower temperatures. Thus, if one could make the system ordered above its melting temperature, it would place the system close to what becomes the global minimum upon cooling. Here, we have shown that a SALR system exposed to steady shear can order above its melting temperature, and thus shearing can be used to facilitate the formation of the lamellar structure.

The analysis of the shear-induced structures we observed was performed after reduction of data dimensionality. We have shown that in the case of such highly anisotropic structures as lamellae, the structural transitions that take place in 3d space can be successfully quantified using PCA applied to 2d binary maps of the structure. The PCA {\color{black}study was supplemented with analysis based on} artificial neural networks of two types: dense and convolutional. Both networks were trained to distinguish between steady states observed at $k_BT/\varepsilon = 0.42$, and their predictions were found to be consistent and in agreement with PCA results.

Moreover, the agreement obtained between the PCA and neural networks analyses suggests that these methods might be combined to form a general framework for accurate machine-learning-based analysis of MD data. In this framework, the coarse position of possible boundaries between structures could first be found with the PCA method. Based on the results of PCA, training data for artificial neural networks would be chosen, and accurate boundaries would be localized by virtue of supervised learning with the neural networks. Further tests with different pattern forming systems are needed to verify the suggested framework.

\section{Acknowledgements}
We would like to acknowledge Jeffrey M. Young for enlightening discussions and comments. We would also like to acknowledge David Nicholson for sharing his implementation of SLLOD in the {\it NPT} ensemble. J.P. acknowledges financial support from the Polish Ministry of Science under the grant Mobility Plus 1631/MOB/V/2017/0. W. R. has received funding from the EU Horizon 2020 programme under the Marie Skłodowska-Curie Grant Agreement No. 665385.

\section{Data Availability Statement}
The data that support the findings of this study are available from the corresponding author
upon reasonable request.


\begin{thebibliography}{53}
\expandafter\ifx\csname natexlab\endcsname\relax\def\natexlab#1{#1}\fi
\expandafter\ifx\csname bibnamefont\endcsname\relax
  \def\bibnamefont#1{#1}\fi
\expandafter\ifx\csname bibfnamefont\endcsname\relax
  \def\bibfnamefont#1{#1}\fi
\expandafter\ifx\csname citenamefont\endcsname\relax
  \def\citenamefont#1{#1}\fi
\expandafter\ifx\csname url\endcsname\relax
  \def\url#1{\texttt{#1}}\fi
\expandafter\ifx\csname urlprefix\endcsname\relax\def\urlprefix{URL }\fi
\providecommand{\bibinfo}[2]{#2}
\providecommand{\eprint}[2][]{\url{#2}}

\bibitem[{\citenamefont{Ghezzi and Earnshaw}(1997)}]{ghezzi:97:0}
\bibinfo{author}{\bibfnamefont{F.}~\bibnamefont{Ghezzi}} \bibnamefont{and}
  \bibinfo{author}{\bibfnamefont{J.}~\bibnamefont{Earnshaw}},
  \bibinfo{journal}{{\it J. Phys.: Condens. Matter}}
  \textbf{\bibinfo{volume}{9}}, \bibinfo{pages}{L517} (\bibinfo{year}{1997}).

\bibitem[{\citenamefont{Stradner et~al.}(2004)\citenamefont{Stradner, Sedgwick,
  Cardinaux, Poon, Egelhaaf, and Schurtenberger}}]{stradner:04:0}
\bibinfo{author}{\bibfnamefont{A.}~\bibnamefont{Stradner}},
  \bibinfo{author}{\bibfnamefont{H.}~\bibnamefont{Sedgwick}},
  \bibinfo{author}{\bibfnamefont{F.}~\bibnamefont{Cardinaux}},
  \bibinfo{author}{\bibfnamefont{W.}~\bibnamefont{Poon}},
  \bibinfo{author}{\bibfnamefont{S.}~\bibnamefont{Egelhaaf}}, \bibnamefont{and}
  \bibinfo{author}{\bibfnamefont{P.}~\bibnamefont{Schurtenberger}},
  \bibinfo{journal}{{\it Nature}} \textbf{\bibinfo{volume}{432}},
  \bibinfo{pages}{492} (\bibinfo{year}{2004}).

\bibitem[{\citenamefont{Sciortino et~al.}(2005)\citenamefont{Sciortino,
  Tartaglia, and Zaccarelli}}]{sciortino:05:0}
\bibinfo{author}{\bibfnamefont{F.}~\bibnamefont{Sciortino}},
  \bibinfo{author}{\bibfnamefont{P.}~\bibnamefont{Tartaglia}},
  \bibnamefont{and}
  \bibinfo{author}{\bibfnamefont{E.}~\bibnamefont{Zaccarelli}},
  \bibinfo{journal}{{\it J. Phys. Chem. B}} \textbf{\bibinfo{volume}{109}},
  \bibinfo{pages}{21942} (\bibinfo{year}{2005}).

\bibitem[{\citenamefont{Toledano et~al.}(2009)\citenamefont{Toledano,
  Sciortino, and Zaccarelli}}]{toledano:09:0}
\bibinfo{author}{\bibfnamefont{J.}~\bibnamefont{Toledano}},
  \bibinfo{author}{\bibfnamefont{F.}~\bibnamefont{Sciortino}},
  \bibnamefont{and}
  \bibinfo{author}{\bibfnamefont{E.}~\bibnamefont{Zaccarelli}},
  \bibinfo{journal}{\textit{Soft Matter}} \textbf{\bibinfo{volume}{5}},
  \bibinfo{pages}{2390} (\bibinfo{year}{2009}).

\bibitem[{\citenamefont{Zhuang et~al.}(2016)\citenamefont{Zhuang, Zhang, and
  Charbonneau}}]{zhuang:16:0}
\bibinfo{author}{\bibfnamefont{Y.}~\bibnamefont{Zhuang}},
  \bibinfo{author}{\bibfnamefont{K.}~\bibnamefont{Zhang}}, \bibnamefont{and}
  \bibinfo{author}{\bibfnamefont{P.}~\bibnamefont{Charbonneau}},
  \bibinfo{journal}{{\it Phys. Rev. Lett.}} \textbf{\bibinfo{volume}{116}},
  \bibinfo{pages}{098301} (\bibinfo{year}{2016}).

\bibitem[{\citenamefont{Zhuang and
  Charbonneau}(2016{\natexlab{a}})}]{zhuang:16:1}
\bibinfo{author}{\bibfnamefont{Y.}~\bibnamefont{Zhuang}} \bibnamefont{and}
  \bibinfo{author}{\bibfnamefont{P.}~\bibnamefont{Charbonneau}},
  \bibinfo{journal}{{\it J. Phys. Chem. B}} \textbf{\bibinfo{volume}{120}},
  \bibinfo{pages}{6178} (\bibinfo{year}{2016}{\natexlab{a}}).

\bibitem[{\citenamefont{Zhuang and
  Charbonneau}(2016{\natexlab{b}})}]{zhuang:16:2}
\bibinfo{author}{\bibfnamefont{Y.}~\bibnamefont{Zhuang}} \bibnamefont{and}
  \bibinfo{author}{\bibfnamefont{P.}~\bibnamefont{Charbonneau}},
  \bibinfo{journal}{{\it J. Phys. Chem. B}} \textbf{\bibinfo{volume}{120}},
  \bibinfo{pages}{7775} (\bibinfo{year}{2016}{\natexlab{b}}).

\bibitem[{\citenamefont{Ciach and G\'o\'zd\'z}(2010)}]{ciach:10:1}
\bibinfo{author}{\bibfnamefont{A.}~\bibnamefont{Ciach}} \bibnamefont{and}
  \bibinfo{author}{\bibfnamefont{W.~T.} \bibnamefont{G\'o\'zd\'z}},
  \bibinfo{journal}{{\it Condens. Matter Phys.}} \textbf{\bibinfo{volume}{13}},
  \bibinfo{pages}{23603} (\bibinfo{year}{2010}).

\bibitem[{\citenamefont{Archer}(2008)}]{archer:08:0}
\bibinfo{author}{\bibfnamefont{A.~J.} \bibnamefont{Archer}},
  \bibinfo{journal}{{\it Phys. Rev. E}} \textbf{\bibinfo{volume}{78}},
  \bibinfo{pages}{031402} (\bibinfo{year}{2008}).

\bibitem[{\citenamefont{Ciach}(2008)}]{ciach:08:1}
\bibinfo{author}{\bibfnamefont{A.}~\bibnamefont{Ciach}}, \bibinfo{journal}{{\it
  Phys. Rev. E}} \textbf{\bibinfo{volume}{78}}, \bibinfo{pages}{061505}
  (\bibinfo{year}{2008}).

\bibitem[{\citenamefont{de~Candia et~al.}(2006)\citenamefont{de~Candia, Gado,
  Fierro, Sator, Tarzia, and Coniglio}}]{candia:06:0}
\bibinfo{author}{\bibfnamefont{A.}~\bibnamefont{de~Candia}},
  \bibinfo{author}{\bibfnamefont{E.~D.} \bibnamefont{Gado}},
  \bibinfo{author}{\bibfnamefont{A.}~\bibnamefont{Fierro}},
  \bibinfo{author}{\bibfnamefont{N.}~\bibnamefont{Sator}},
  \bibinfo{author}{\bibfnamefont{M.}~\bibnamefont{Tarzia}}, \bibnamefont{and}
  \bibinfo{author}{\bibfnamefont{A.}~\bibnamefont{Coniglio}},
  \bibinfo{journal}{{\it Phys. Rev. E}} \textbf{\bibinfo{volume}{74}},
  \bibinfo{pages}{010403(R)} (\bibinfo{year}{2006}).

\bibitem[{\citenamefont{Law et~al.}(2013)\citenamefont{Law, Auriol, Smith,
  Horozov, and Buzza}}]{law:13:0}
\bibinfo{author}{\bibfnamefont{A.~D.} \bibnamefont{Law}},
  \bibinfo{author}{\bibfnamefont{M.}~\bibnamefont{Auriol}},
  \bibinfo{author}{\bibfnamefont{D.}~\bibnamefont{Smith}},
  \bibinfo{author}{\bibfnamefont{T.~S.} \bibnamefont{Horozov}},
  \bibnamefont{and} \bibinfo{author}{\bibfnamefont{D.~M.~A.}
  \bibnamefont{Buzza}}, \bibinfo{journal}{{\it Phys. Rev. Lett.}}
  \textbf{\bibinfo{volume}{110}}, \bibinfo{pages}{138301}
  (\bibinfo{year}{2013}).

\bibitem[{\citenamefont{Sear and Gelbart}(1999)}]{sear:99:0}
\bibinfo{author}{\bibfnamefont{R.~P.} \bibnamefont{Sear}} \bibnamefont{and}
  \bibinfo{author}{\bibfnamefont{W.~M.} \bibnamefont{Gelbart}},
  \bibinfo{journal}{{\it J. Chem. Phys.}} \textbf{\bibinfo{volume}{110}},
  \bibinfo{pages}{4582} (\bibinfo{year}{1999}).

\bibitem[{\citenamefont{Royall}(2018)}]{royall:18:0}
\bibinfo{author}{\bibfnamefont{C.~P.} \bibnamefont{Royall}},
  \bibinfo{journal}{Soft Matter} \textbf{\bibinfo{volume}{14}},
  \bibinfo{pages}{4020} (\bibinfo{year}{2018}).

\bibitem[{\citenamefont{Almarza et~al.}(2016)\citenamefont{Almarza,
  P\c{e}kalski, and Ciach}}]{almarza:16:0}
\bibinfo{author}{\bibfnamefont{N.~G.} \bibnamefont{Almarza}},
  \bibinfo{author}{\bibfnamefont{J.}~\bibnamefont{P\c{e}kalski}},
  \bibnamefont{and} \bibinfo{author}{\bibfnamefont{A.}~\bibnamefont{Ciach}},
  \bibinfo{journal}{\textit{Soft Matter}} \textbf{\bibinfo{volume}{12}},
  \bibinfo{pages}{7551} (\bibinfo{year}{2016}).

\bibitem[{\citenamefont{P{\c e}kalski et~al.}(2019)\citenamefont{P{\c e}kalski,
  Bildanau, and Ciach}}]{pekalski:19:0}
\bibinfo{author}{\bibfnamefont{J.}~\bibnamefont{P{\c e}kalski}},
  \bibinfo{author}{\bibfnamefont{E.}~\bibnamefont{Bildanau}}, \bibnamefont{and}
  \bibinfo{author}{\bibfnamefont{A.}~\bibnamefont{Ciach}},
  \bibinfo{journal}{{\it Soft Matter}} \textbf{\bibinfo{volume}{15}},
  \bibinfo{pages}{7715} (\bibinfo{year}{2019}),
  \urlprefix\url{http://dx.doi.org/10.1039/C9SM01179J}.

\bibitem[{\citenamefont{Palberg et~al.}(1995)\citenamefont{Palberg, M{\"o}nch,
  Schwarz, and Leiderer}}]{palberg:95:0}
\bibinfo{author}{\bibfnamefont{T.}~\bibnamefont{Palberg}},
  \bibinfo{author}{\bibfnamefont{W.}~\bibnamefont{M{\"o}nch}},
  \bibinfo{author}{\bibfnamefont{J.}~\bibnamefont{Schwarz}}, \bibnamefont{and}
  \bibinfo{author}{\bibfnamefont{P.}~\bibnamefont{Leiderer}},
  \bibinfo{journal}{{\it J. Chem. Phys.}} \textbf{\bibinfo{volume}{102}},
  \bibinfo{pages}{5082} (\bibinfo{year}{1995}).

\bibitem[{\citenamefont{Blaak et~al.}(2004)\citenamefont{Blaak, Auer, Frenkel,
  and L{\"o}wen}}]{blaak:04:0}
\bibinfo{author}{\bibfnamefont{R.}~\bibnamefont{Blaak}},
  \bibinfo{author}{\bibfnamefont{S.}~\bibnamefont{Auer}},
  \bibinfo{author}{\bibfnamefont{D.}~\bibnamefont{Frenkel}}, \bibnamefont{and}
  \bibinfo{author}{\bibfnamefont{H.}~\bibnamefont{L{\"o}wen}},
  \bibinfo{journal}{{\it Phys. Rev. Lett.}} \textbf{\bibinfo{volume}{93}},
  \bibinfo{pages}{068303} (\bibinfo{year}{2004}).

\bibitem[{\citenamefont{Ackerson and Pusey}(1988)}]{ackerson:88:0}
\bibinfo{author}{\bibfnamefont{B.~J.} \bibnamefont{Ackerson}} \bibnamefont{and}
  \bibinfo{author}{\bibfnamefont{P.}~\bibnamefont{Pusey}},
  \bibinfo{journal}{{\it Phys. Rev. Lett.}} \textbf{\bibinfo{volume}{61}},
  \bibinfo{pages}{1033} (\bibinfo{year}{1988}).

\bibitem[{\citenamefont{Amos et~al.}(2000)\citenamefont{Amos, Rarity, Tapster,
  Shepherd, and Kitson}}]{amos:00:0}
\bibinfo{author}{\bibfnamefont{R.}~\bibnamefont{Amos}},
  \bibinfo{author}{\bibfnamefont{J.}~\bibnamefont{Rarity}},
  \bibinfo{author}{\bibfnamefont{P.}~\bibnamefont{Tapster}},
  \bibinfo{author}{\bibfnamefont{T.}~\bibnamefont{Shepherd}}, \bibnamefont{and}
  \bibinfo{author}{\bibfnamefont{S.}~\bibnamefont{Kitson}},
  \bibinfo{journal}{{\it Phys. Rev. E}} \textbf{\bibinfo{volume}{61}},
  \bibinfo{pages}{2929} (\bibinfo{year}{2000}).

\bibitem[{\citenamefont{Catherall et~al.}(2000)\citenamefont{Catherall,
  Melrose, and Ball}}]{catherall:00:0}
\bibinfo{author}{\bibfnamefont{A.~A.} \bibnamefont{Catherall}},
  \bibinfo{author}{\bibfnamefont{J.~R.} \bibnamefont{Melrose}},
  \bibnamefont{and} \bibinfo{author}{\bibfnamefont{R.~C.} \bibnamefont{Ball}},
  \bibinfo{journal}{{\it J. Rheol.}} \textbf{\bibinfo{volume}{44}},
  \bibinfo{pages}{1} (\bibinfo{year}{2000}).

\bibitem[{\citenamefont{Li and de~Jeu}(2004)}]{li:04:0}
\bibinfo{author}{\bibfnamefont{L.}~\bibnamefont{Li}} \bibnamefont{and}
  \bibinfo{author}{\bibfnamefont{W.~H.} \bibnamefont{de~Jeu}},
  \bibinfo{journal}{{\it Phys. Rev. Lett.}} \textbf{\bibinfo{volume}{92}},
  \bibinfo{pages}{075506} (\bibinfo{year}{2004}).

\bibitem[{\citenamefont{Harrison et~al.}(2000)\citenamefont{Harrison, Adamson,
  Cheng, Sebastian, Sethuraman, Huse, Register, and Chaikin}}]{harrison:20:0}
\bibinfo{author}{\bibfnamefont{C.}~\bibnamefont{Harrison}},
  \bibinfo{author}{\bibfnamefont{D.~H.} \bibnamefont{Adamson}},
  \bibinfo{author}{\bibfnamefont{Z.}~\bibnamefont{Cheng}},
  \bibinfo{author}{\bibfnamefont{J.~M.} \bibnamefont{Sebastian}},
  \bibinfo{author}{\bibfnamefont{S.}~\bibnamefont{Sethuraman}},
  \bibinfo{author}{\bibfnamefont{D.~A.} \bibnamefont{Huse}},
  \bibinfo{author}{\bibfnamefont{R.~A.} \bibnamefont{Register}},
  \bibnamefont{and} \bibinfo{author}{\bibfnamefont{P.~M.}
  \bibnamefont{Chaikin}}, \bibinfo{journal}{{\it Science}}
  \textbf{\bibinfo{volume}{290}}, \bibinfo{pages}{1558} (\bibinfo{year}{2000}).

\bibitem[{\citenamefont{Nikoubashman et~al.}(2014)\citenamefont{Nikoubashman,
  Davis, Michal, Chaikin, Register, and Panagiotopoulos}}]{nikoubashman:14:0}
\bibinfo{author}{\bibfnamefont{A.}~\bibnamefont{Nikoubashman}},
  \bibinfo{author}{\bibfnamefont{R.~L.} \bibnamefont{Davis}},
  \bibinfo{author}{\bibfnamefont{B.~T.} \bibnamefont{Michal}},
  \bibinfo{author}{\bibfnamefont{P.~M.} \bibnamefont{Chaikin}},
  \bibinfo{author}{\bibfnamefont{R.~A.} \bibnamefont{Register}},
  \bibnamefont{and} \bibinfo{author}{\bibfnamefont{A.~Z.}
  \bibnamefont{Panagiotopoulos}}, \bibinfo{journal}{{\it ACS nano}}
  \textbf{\bibinfo{volume}{8}}, \bibinfo{pages}{8015} (\bibinfo{year}{2014}).

\bibitem[{\citenamefont{Imperio et~al.}(2008)\citenamefont{Imperio, Reatto, and
  Zapperi}}]{imperio:08:0}
\bibinfo{author}{\bibfnamefont{A.}~\bibnamefont{Imperio}},
  \bibinfo{author}{\bibfnamefont{L.}~\bibnamefont{Reatto}}, \bibnamefont{and}
  \bibinfo{author}{\bibfnamefont{S.}~\bibnamefont{Zapperi}},
  \bibinfo{journal}{{\it Phys. Rev. E}} \textbf{\bibinfo{volume}{78}},
  \bibinfo{pages}{021402} (\bibinfo{year}{2008}).

\bibitem[{\citenamefont{Ruiz-Franco et~al.}(2019)\citenamefont{Ruiz-Franco,
  Gnan, and Zaccarelli}}]{ruiz:19:0}
\bibinfo{author}{\bibfnamefont{J.}~\bibnamefont{Ruiz-Franco}},
  \bibinfo{author}{\bibfnamefont{N.}~\bibnamefont{Gnan}}, \bibnamefont{and}
  \bibinfo{author}{\bibfnamefont{E.}~\bibnamefont{Zaccarelli}},
  \bibinfo{journal}{{\it J. Chem. Phys.}} \textbf{\bibinfo{volume}{150}},
  \bibinfo{pages}{024905} (\bibinfo{year}{2019}).

\bibitem[{\citenamefont{Stopper and Roth}(2018)}]{stopper:18:0}
\bibinfo{author}{\bibfnamefont{D.}~\bibnamefont{Stopper}} \bibnamefont{and}
  \bibinfo{author}{\bibfnamefont{R.}~\bibnamefont{Roth}},
  \bibinfo{journal}{{\it Phys. Rev. E}} \textbf{\bibinfo{volume}{97}},
  \bibinfo{pages}{062602} (\bibinfo{year}{2018}).

\bibitem[{\citenamefont{Ciach et~al.}(2013)\citenamefont{Ciach, P{\c e}kalski,
  and G\'o\'zd\'z}}]{ciach:13:0}
\bibinfo{author}{\bibfnamefont{A.}~\bibnamefont{Ciach}},
  \bibinfo{author}{\bibfnamefont{J.}~\bibnamefont{P{\c e}kalski}},
  \bibnamefont{and} \bibinfo{author}{\bibfnamefont{W.~T.}
  \bibnamefont{G\'o\'zd\'z}}, \bibinfo{journal}{\textit{Soft Matter}}
  \textbf{\bibinfo{volume}{9}}, \bibinfo{pages}{6301} (\bibinfo{year}{2013}).

\bibitem[{\citenamefont{Almarza et~al.}(2014)\citenamefont{Almarza, P{\c
  e}kalski, and Ciach}}]{almarza:14:0}
\bibinfo{author}{\bibfnamefont{N.~G.} \bibnamefont{Almarza}},
  \bibinfo{author}{\bibfnamefont{J.}~\bibnamefont{P{\c e}kalski}},
  \bibnamefont{and} \bibinfo{author}{\bibfnamefont{A.}~\bibnamefont{Ciach}},
  \bibinfo{journal}{{\it J. Chem. Phys.}} \textbf{\bibinfo{volume}{140}},
  \bibinfo{pages}{164708} (\bibinfo{year}{2014}).

\bibitem[{\citenamefont{P{\c e}kalski et~al.}(2013)\citenamefont{P{\c e}kalski,
  Ciach, and Almarza}}]{pekalski:13:0}
\bibinfo{author}{\bibfnamefont{J.}~\bibnamefont{P{\c e}kalski}},
  \bibinfo{author}{\bibfnamefont{A.}~\bibnamefont{Ciach}}, \bibnamefont{and}
  \bibinfo{author}{\bibfnamefont{N.~G.} \bibnamefont{Almarza}},
  \bibinfo{journal}{{\it J. Chem. Phys.}} \textbf{\bibinfo{volume}{138}},
  \bibinfo{pages}{144903} (\bibinfo{year}{2013}).

\bibitem[{\citenamefont{Imperio and Reatto}(2004)}]{imperio:04:0}
\bibinfo{author}{\bibfnamefont{A.}~\bibnamefont{Imperio}} \bibnamefont{and}
  \bibinfo{author}{\bibfnamefont{L.}~\bibnamefont{Reatto}},
  \bibinfo{journal}{{\it J. Phys.: Condens. Matter}}
  \textbf{\bibinfo{volume}{18}}, \bibinfo{pages}{S3769} (\bibinfo{year}{2004}).

\bibitem[{\citenamefont{Carleo et~al.}(2019)\citenamefont{Carleo, Cirac,
  Cranmer, Daudet, Schuld, Tishby, Vogt-Maranto, and
  Zdeborov\'a}}]{carleo:19:0}
\bibinfo{author}{\bibfnamefont{G.}~\bibnamefont{Carleo}},
  \bibinfo{author}{\bibfnamefont{I.}~\bibnamefont{Cirac}},
  \bibinfo{author}{\bibfnamefont{K.}~\bibnamefont{Cranmer}},
  \bibinfo{author}{\bibfnamefont{L.}~\bibnamefont{Daudet}},
  \bibinfo{author}{\bibfnamefont{M.}~\bibnamefont{Schuld}},
  \bibinfo{author}{\bibfnamefont{N.}~\bibnamefont{Tishby}},
  \bibinfo{author}{\bibfnamefont{L.}~\bibnamefont{Vogt-Maranto}},
  \bibnamefont{and}
  \bibinfo{author}{\bibfnamefont{L.}~\bibnamefont{Zdeborov\'a}},
  \bibinfo{journal}{Rev. Mod. Phys.} \textbf{\bibinfo{volume}{91}},
  \bibinfo{pages}{045002} (\bibinfo{year}{2019}),
  \urlprefix\url{https://link.aps.org/doi/10.1103/RevModPhys.91.045002}.

\bibitem[{\citenamefont{Mehta et~al.}(2019)\citenamefont{Mehta, Bukov, Wang,
  Day, Richardson, Fisher, and Schwab}}]{mehta:19:0}
\bibinfo{author}{\bibfnamefont{P.}~\bibnamefont{Mehta}},
  \bibinfo{author}{\bibfnamefont{M.}~\bibnamefont{Bukov}},
  \bibinfo{author}{\bibfnamefont{C.-H.} \bibnamefont{Wang}},
  \bibinfo{author}{\bibfnamefont{A.~G.} \bibnamefont{Day}},
  \bibinfo{author}{\bibfnamefont{C.}~\bibnamefont{Richardson}},
  \bibinfo{author}{\bibfnamefont{C.~K.} \bibnamefont{Fisher}},
  \bibnamefont{and} \bibinfo{author}{\bibfnamefont{D.~J.}
  \bibnamefont{Schwab}}, \bibinfo{journal}{Physics reports}
  (\bibinfo{year}{2019}).

\bibitem[{\citenamefont{Carrasquilla and Melko}(2017)}]{carrasquilla:17:0}
\bibinfo{author}{\bibfnamefont{J.}~\bibnamefont{Carrasquilla}}
  \bibnamefont{and} \bibinfo{author}{\bibfnamefont{R.~G.} \bibnamefont{Melko}},
  \bibinfo{journal}{Nature Physics} \textbf{\bibinfo{volume}{13}},
  \bibinfo{pages}{431} (\bibinfo{year}{2017}).

\bibitem[{\citenamefont{Sciortino et~al.}(2004)\citenamefont{Sciortino, Mossa,
  Zaccarelli, and Tartaglia}}]{sciortino:04:0}
\bibinfo{author}{\bibfnamefont{F.}~\bibnamefont{Sciortino}},
  \bibinfo{author}{\bibfnamefont{S.}~\bibnamefont{Mossa}},
  \bibinfo{author}{\bibfnamefont{E.}~\bibnamefont{Zaccarelli}},
  \bibnamefont{and}
  \bibinfo{author}{\bibfnamefont{P.}~\bibnamefont{Tartaglia}},
  \bibinfo{journal}{{\it Phys. Rev. Lett.}} \textbf{\bibinfo{volume}{93}},
  \bibinfo{pages}{055701} (\bibinfo{year}{2004}).

\bibitem[{\citenamefont{Mani et~al.}(2014)\citenamefont{Mani, Lechner, Kegel,
  and Bolhuis}}]{mani:14:0}
\bibinfo{author}{\bibfnamefont{E.}~\bibnamefont{Mani}},
  \bibinfo{author}{\bibfnamefont{W.}~\bibnamefont{Lechner}},
  \bibinfo{author}{\bibfnamefont{W.~K.} \bibnamefont{Kegel}}, \bibnamefont{and}
  \bibinfo{author}{\bibfnamefont{P.~G.} \bibnamefont{Bolhuis}},
  \bibinfo{journal}{\textit{Soft Matter}} \textbf{\bibinfo{volume}{10}},
  \bibinfo{pages}{4479} (\bibinfo{year}{2014}),
  \urlprefix\url{http://dx.doi.org/10.1039/C3SM53058B}.

\bibitem[{\citenamefont{Santos et~al.}(2017)\citenamefont{Santos, P\c{e}kalski,
  and Panagiotopoulos}}]{santos:17:0}
\bibinfo{author}{\bibfnamefont{A.~P.} \bibnamefont{Santos}},
  \bibinfo{author}{\bibfnamefont{J.}~\bibnamefont{P\c{e}kalski}},
  \bibnamefont{and} \bibinfo{author}{\bibfnamefont{A.~Z.}
  \bibnamefont{Panagiotopoulos}}, \bibinfo{journal}{{\it Soft Matter}}
  \textbf{\bibinfo{volume}{13}}, \bibinfo{pages}{8055} (\bibinfo{year}{2017}).

\bibitem[{\citenamefont{Glaser et~al.}(2015)\citenamefont{Glaser, Nguyen,
  Anderson, Lui, Spiga, Millan, Morse, and Glotzer}}]{hoomd:2}
\bibinfo{author}{\bibfnamefont{J.}~\bibnamefont{Glaser}},
  \bibinfo{author}{\bibfnamefont{T.~D.} \bibnamefont{Nguyen}},
  \bibinfo{author}{\bibfnamefont{J.~A.} \bibnamefont{Anderson}},
  \bibinfo{author}{\bibfnamefont{P.}~\bibnamefont{Lui}},
  \bibinfo{author}{\bibfnamefont{F.}~\bibnamefont{Spiga}},
  \bibinfo{author}{\bibfnamefont{J.~A.} \bibnamefont{Millan}},
  \bibinfo{author}{\bibfnamefont{D.~C.} \bibnamefont{Morse}}, \bibnamefont{and}
  \bibinfo{author}{\bibfnamefont{S.~C.} \bibnamefont{Glotzer}},
  \bibinfo{journal}{\textit{Comput. Phys. Commun.}}
  \textbf{\bibinfo{volume}{192}}, \bibinfo{pages}{97} (\bibinfo{year}{2015}).

\bibitem[{\citenamefont{Anderson et~al.}(2008)\citenamefont{Anderson, Lorenz,
  and Travesset}}]{hoomd:1}
\bibinfo{author}{\bibfnamefont{J.~A.} \bibnamefont{Anderson}},
  \bibinfo{author}{\bibfnamefont{C.~D.} \bibnamefont{Lorenz}},
  \bibnamefont{and}
  \bibinfo{author}{\bibfnamefont{A.}~\bibnamefont{Travesset}},
  \bibinfo{journal}{\textit{J. Comput. Phys.}} \textbf{\bibinfo{volume}{227}},
  \bibinfo{pages}{5342} (\bibinfo{year}{2008}).

\bibitem[{\citenamefont{Plimpton}(1995)}]{lammps:1}
\bibinfo{author}{\bibfnamefont{S.}~\bibnamefont{Plimpton}},
  \bibinfo{journal}{{\it J. Comput. Phys.}} \textbf{\bibinfo{volume}{117}},
  \bibinfo{pages}{1} (\bibinfo{year}{1995}),
  \urlprefix\url{http://lammps.sandia.gov}.

\bibitem[{\citenamefont{Todd and Daivis}(2017)}]{sllod:book}
\bibinfo{author}{\bibfnamefont{B.~D.} \bibnamefont{Todd}} \bibnamefont{and}
  \bibinfo{author}{\bibfnamefont{P.~J.} \bibnamefont{Daivis}},
  \emph{\bibinfo{title}{Nonequilibrium molecular dynamics: theory, algorithms
  and applications}} (\bibinfo{publisher}{Cambridge University Press},
  \bibinfo{year}{2017}).

\bibitem[{\citenamefont{Jolliffe and Cadima}(2016)}]{jolliffe:16:0}
\bibinfo{author}{\bibfnamefont{I.~T.} \bibnamefont{Jolliffe}} \bibnamefont{and}
  \bibinfo{author}{\bibfnamefont{J.}~\bibnamefont{Cadima}},
  \bibinfo{journal}{{\it Philos. Trans. R. Soc. A}}
  \textbf{\bibinfo{volume}{374}}, \bibinfo{pages}{20150202}
  (\bibinfo{year}{2016}).

\bibitem[{\citenamefont{Wetzel}(2017)}]{wetzel:17:0}
\bibinfo{author}{\bibfnamefont{S.~J.} \bibnamefont{Wetzel}},
  \bibinfo{journal}{{\it Phys. Rev. E}} \textbf{\bibinfo{volume}{96}},
  \bibinfo{pages}{022140} (\bibinfo{year}{2017}),
  \urlprefix\url{https://link.aps.org/doi/10.1103/PhysRevE.96.022140}.

\bibitem[{\citenamefont{Cubuk et~al.}(2015)\citenamefont{Cubuk, Schoenholz,
  Rieser, Malone, Rottler, Durian, Kaxiras, and Liu}}]{cubuk:15:0}
\bibinfo{author}{\bibfnamefont{E.~D.} \bibnamefont{Cubuk}},
  \bibinfo{author}{\bibfnamefont{S.~S.} \bibnamefont{Schoenholz}},
  \bibinfo{author}{\bibfnamefont{J.~M.} \bibnamefont{Rieser}},
  \bibinfo{author}{\bibfnamefont{B.~D.} \bibnamefont{Malone}},
  \bibinfo{author}{\bibfnamefont{J.}~\bibnamefont{Rottler}},
  \bibinfo{author}{\bibfnamefont{D.~J.} \bibnamefont{Durian}},
  \bibinfo{author}{\bibfnamefont{E.}~\bibnamefont{Kaxiras}}, \bibnamefont{and}
  \bibinfo{author}{\bibfnamefont{A.~J.} \bibnamefont{Liu}},
  \bibinfo{journal}{{\it Phys. Rev. Lett.}} \textbf{\bibinfo{volume}{114}},
  \bibinfo{pages}{108001} (\bibinfo{year}{2015}).

\bibitem[{\citenamefont{Schoenholz et~al.}(2016)\citenamefont{Schoenholz,
  Cubuk, Sussman, Kaxiras, and Liu}}]{schoenholz:16:0}
\bibinfo{author}{\bibfnamefont{S.~S.} \bibnamefont{Schoenholz}},
  \bibinfo{author}{\bibfnamefont{E.~D.} \bibnamefont{Cubuk}},
  \bibinfo{author}{\bibfnamefont{D.~M.} \bibnamefont{Sussman}},
  \bibinfo{author}{\bibfnamefont{E.}~\bibnamefont{Kaxiras}}, \bibnamefont{and}
  \bibinfo{author}{\bibfnamefont{A.~J.} \bibnamefont{Liu}},
  \bibinfo{journal}{{\it Nat. Phys.}} \textbf{\bibinfo{volume}{12}},
  \bibinfo{pages}{469} (\bibinfo{year}{2016}).

\bibitem[{\citenamefont{Schoenholz et~al.}(2017)\citenamefont{Schoenholz,
  Cubuk, Kaxiras, and Liu}}]{schoenholz:17:0}
\bibinfo{author}{\bibfnamefont{S.~S.} \bibnamefont{Schoenholz}},
  \bibinfo{author}{\bibfnamefont{E.~D.} \bibnamefont{Cubuk}},
  \bibinfo{author}{\bibfnamefont{E.}~\bibnamefont{Kaxiras}}, \bibnamefont{and}
  \bibinfo{author}{\bibfnamefont{A.~J.} \bibnamefont{Liu}},
  \bibinfo{journal}{{\it Proc. Nat. Acad. Sci. USA}}
  \textbf{\bibinfo{volume}{114}}, \bibinfo{pages}{263} (\bibinfo{year}{2017}),
  ISSN \bibinfo{issn}{0027-8424},
  \eprint{https://www.pnas.org/content/114/2/263.full.pdf},
  \urlprefix\url{https://www.pnas.org/content/114/2/263}.

\bibitem[{\citenamefont{Cristoforetti et~al.}(2017)\citenamefont{Cristoforetti,
  Jurman, Nardelli, and Furlanello}}]{cristoforetti:17:0}
\bibinfo{author}{\bibfnamefont{M.}~\bibnamefont{Cristoforetti}},
  \bibinfo{author}{\bibfnamefont{G.}~\bibnamefont{Jurman}},
  \bibinfo{author}{\bibfnamefont{A.~I.} \bibnamefont{Nardelli}},
  \bibnamefont{and}
  \bibinfo{author}{\bibfnamefont{C.}~\bibnamefont{Furlanello}},
  \bibinfo{journal}{arXiv preprint arXiv:1705.09524}  (\bibinfo{year}{2017}).

\bibitem[{\citenamefont{Daniel}(2013)}]{graupe:13:0}
\bibinfo{author}{\bibfnamefont{G.}~\bibnamefont{Daniel}},
  \emph{\bibinfo{title}{Principles of artificial neural networks}},
  vol.~\bibinfo{volume}{7} (\bibinfo{publisher}{World Scientific},
  \bibinfo{year}{2013}).

\bibitem[{\citenamefont{Gu et~al.}(2018)\citenamefont{Gu, Wang, Kuen, Ma,
  Shahroudy, Shuai, Liu, Wang, Wang, Cai et~al.}}]{gu:18:0}
\bibinfo{author}{\bibfnamefont{J.}~\bibnamefont{Gu}},
  \bibinfo{author}{\bibfnamefont{Z.}~\bibnamefont{Wang}},
  \bibinfo{author}{\bibfnamefont{J.}~\bibnamefont{Kuen}},
  \bibinfo{author}{\bibfnamefont{L.}~\bibnamefont{Ma}},
  \bibinfo{author}{\bibfnamefont{A.}~\bibnamefont{Shahroudy}},
  \bibinfo{author}{\bibfnamefont{B.}~\bibnamefont{Shuai}},
  \bibinfo{author}{\bibfnamefont{T.}~\bibnamefont{Liu}},
  \bibinfo{author}{\bibfnamefont{X.}~\bibnamefont{Wang}},
  \bibinfo{author}{\bibfnamefont{G.}~\bibnamefont{Wang}},
  \bibinfo{author}{\bibfnamefont{J.}~\bibnamefont{Cai}}, \bibnamefont{et~al.},
  \bibinfo{journal}{{\it Pattern Recognit.}} \textbf{\bibinfo{volume}{77}},
  \bibinfo{pages}{354} (\bibinfo{year}{2018}).

\bibitem[{\citenamefont{Dhillon and Verma}(2019)}]{dhillon:19:0}
\bibinfo{author}{\bibfnamefont{A.}~\bibnamefont{Dhillon}} \bibnamefont{and}
  \bibinfo{author}{\bibfnamefont{G.~K.} \bibnamefont{Verma}},
  \bibinfo{journal}{{{\it Prog. Artif. Intell.}}}  (\bibinfo{year}{2019}), ISSN
  \bibinfo{issn}{2192-6360},
  \urlprefix\url{https://doi.org/10.1007/s13748-019-00203-0}.

\bibitem[{\citenamefont{Nair and Hinton}(2010)}]{nair:10:0}
\bibinfo{author}{\bibfnamefont{V.}~\bibnamefont{Nair}} \bibnamefont{and}
  \bibinfo{author}{\bibfnamefont{G.~E.} \bibnamefont{Hinton}}, in
  \emph{\bibinfo{booktitle}{Proceedings of the 27th international conference on
  machine learning (ICML-10)}} (\bibinfo{year}{2010}), pp.
  \bibinfo{pages}{807--814}.

\bibitem[{\citenamefont{Stukowski}(2010)}]{ovito}
\bibinfo{author}{\bibfnamefont{A.}~\bibnamefont{Stukowski}},
  \bibinfo{journal}{{\it Modell. Simul. Mater. Sci. Eng.}}
  \textbf{\bibinfo{volume}{18}}, \bibinfo{pages}{015012}
  (\bibinfo{year}{2010}),
  \urlprefix\url{http://stacks.iop.org/0965-0393/18/i=1/a=015012}.

\bibitem[{\citenamefont{Olson~Reichhardt
  et~al.}(2010)\citenamefont{Olson~Reichhardt, Reichhardt, and
  Bishop}}]{Reichhardt2010}
\bibinfo{author}{\bibfnamefont{C.~J.} \bibnamefont{Olson~Reichhardt}},
  \bibinfo{author}{\bibfnamefont{C.}~\bibnamefont{Reichhardt}},
  \bibnamefont{and} \bibinfo{author}{\bibfnamefont{A.~R.}
  \bibnamefont{Bishop}}, \bibinfo{journal}{{\it Phys. Rev. E}}
  \textbf{\bibinfo{volume}{82}}, \bibinfo{pages}{041502}
  (\bibinfo{year}{2010}),
  \urlprefix\url{http://link.aps.org/doi/10.1103/PhysRevE.82.041502}.

\end{thebibliography}
\end{document}